\documentclass[aps]{revtex4}
\usepackage{graphicx}
\begin{document}
\title{Stress variations near surfaces in diamond-like amorphous carbon}
\author{M. Fyta$^{(1)}$ and P. C. Kelires$^{(1,2)}$}
\affiliation{$^{(1)}$Physics Department, University of Crete, P.O. Box 2208,
710 03 Heraclion, Crete, Greece \\
$^{(2)}$Foundation for Research and Technology-Hellas (FORTH),
P.O. Box 1527, 711 10, Heraklion, Crete, Greece.}

\begin{abstract}
Using Monte Carlo simulations within the empirical potential approach,
we examine the effect produced by the surface environment on the
atomic level stresses in tetrahedral amorphous carbon. Both the distribution
of stresses and the distributions of sp$^{2}$ and sp$^{3}$ atoms 
as a function of depth from the surface are highly inhomogeneous. They 
show the same close relationship between local stress and bonding 
hybridization found previously in the bulk of the material. Compressive
local stress favors the formation of sp$^{3}$ sites, while tensile
stress favors the formation of sp$^{2}$ sites.
\end{abstract}
\maketitle

\section{Introduction}

One of the important properties of diamond-like tetrahedral amorphous
carbon (ta-C), is believed to be its possession of high 
average compressive stress. The presence of such intrinsic stress is
desirable, in one aspect, because it is associated with a high
percentage of tetrahedral (sp$^{3}$) atoms. However, it is also desirable
to relieve the compressive stress because is believed to produce
adhesion failure at the film/substrate interface. The origin of this
stress and its relationship to the formation of sp$^{3}$ sites is
still a matter of debate \cite{McKenz1}. One view is that the
compressive stress is produced by the energy of ion bombardment in the
deposition process, and that it is responsible for a thermodynamic
phase transition from sp$^{2}$ $\rightarrow$ sp$^{3}$ phase at a
critical average value \cite{McKenz2}. A model to explain the dependence
of compressive stress on ion energy has been developed \cite{Davis1}.
Another view is that ``subplantation'' events (i.e., shallow implantation
of incoming atoms) during deposition produce densification, and as a 
result this promotes the sp$^{3}$ bonding and the generation of compressive 
stress \cite{Rober1}.

An alternative approach to this issue is based on the concept
of {\it atomic level stresses} \cite{Kelires1}. The logic behind it rests
on the observation that stress generation is a local process extending
over few atomic volumes, and that what really matters is 
the local character of the stress tensor.
It was shown that the {\it average} compressive stress
is not a crucial factor for the stabilization of sp$^{3}$ sites, as it
could be negligibly small \cite{Kelires1}. Note that current experimental 
efforts to solve the adhesion problem \cite{Friedmann,Ferrari} aim to
relieve the average stress through post-growth thermal annealing.
Instead, it is proposed that the compressive local stresses are the
crucial parameter. In this approach, bonding hybridization is closely
associated with the local stress conditions: compressive stresses
favor the formation of sp$^{3}$ sites, while tensile conditions favor
sp$^{2}$ sites. This correspondence was derived from considerations
of the bulk material.

In this paper, we take into account the surface environment and
examine how global this relationship is. By analyzing consistently
the variations of stresses and the distributions of atoms as functions
of depth, we unambigiously show the validity of the local stress - 
bonding hybridization relationship in ta-C.
In the following, we first outline the method used in
our calculations, and then we give the results regarding the surface 
microstructure and atomic stresses, and discuss 
their association.

\section{Theoretical methods}

The investigations are based on Monte Carlo (MC) simulations within the
empirical potential approach \cite{Kelires1}. 
We use slab supercells with two free
surfaces, and with periodic boundary conditions in the two lateral 
dimensions. The cells are deep enough ($\sim$ 30 \AA\ in each direction)
to sufficiently describe the variations with depth. They contain
720 atoms.

The amorphous networks are generated by quenching the liquid under
pressure within the constant pressure-temperature (N,P,T) ensemble.
The liquid is equilibrated at $\sim$ 9000 K and then cooled to 300 K 
under various pressures up to 300 GPa and at cooling rates up to $\sim$ 25 
(MC steps)/atom-K. Alternatively, one can use the constant-volume
(N,V,T) ensemble and reflecting walls at the two surfaces to contain the
liquid while quenching. After removing the pressure (or the walls), 
the cell density is equilibrated and the atomic positions are 
thoroughly relaxed.

The empirical potential of Tersoff \cite{Tersoff1} is used
to model the interactions. It is extensively tested and found to
describe reasonably well the properties of ta-C \cite{Kelires1,Kelires2}. 
The potential describes strain quite well \cite{Kelires3}.
Though in principle less accurate than {\it ab initio} methods, this 
approach permits us to extract the {\it atomic level stresses}, which are
otherwise inaccessible. The atomic stresses are in general the result 
of local incompatibilities, usually due to atomic size mismatch. 
They are defined \cite{Kelires1} by considering an atomic compression (tension)
\begin{equation}
\sigma_{i}=-dE_{i}/dlnV \sim p_{i}\Omega_{i},
\label{astress}
\end{equation}
where $E_{i}$ is the energy of atom $i$ and $V$ is the volume. 
Note that the decomposition of total energy into atomic contributions is not
possible using first-principles methods. Dividing
by the atomic volume $\Omega_{i}$ converts $\sigma_{i}$ 
into units of pressure $p_{i}$.
This local hydrostatic pressure describes the local density fluctuations.
The total intrinsic stress of the system can be calculated by summing up
the $\sigma_{i}$ over all atoms. For a completely strain compensated
system the total stress is zero. This means that the individual
contributions cancel each other, but it does not mean that themselves are
also diminished.

\section{Results}

\subsection{ {\it Structure} }

We first briefly describe the surface structure of ta-C which has been
simulated with the same MC methodology earlier \cite{Kelires4}. 
It was found that the amorphous surface 
layer reconstructs to form distorted sixfold-ring patterns 
of sp$^{2}$ sites (along with odd-membered rings and fourfold atoms)
in a graphite-like manner. This is illustrated here in Fig. 1, showing 
the structure of one of the two surfaces in a typical slab cell with 
average coordination $\bar{z} \simeq$ 3.75. It seems that this structure is
a generic feature of ta-C, supported also by other recent 
simulations \cite{Dong}.
This specific feature of the surface layer has not yet been identified
by experiment, although the enhancement with sp$^{2}$ sites is
established \cite{Davis2}. The graphite-like
nature of the surface in ta-C is in sharp contrast to its bulk 
topology, for which researchers have reached a consensus about the 
clustering of sp$^{2}$ sites: they are found in pairs, chains and small
clusters rather than in aromatic rings.

It is essential for our purposes to have a detailed picture of how
sp$^{2}$ and sp$^{3}$ atoms are distributed in the slab cells as a
function of distance z from the two surfaces. We thus compute the
probabilities (unnormalized) of finding threefold and fourfold atoms
at depth z in the slab cell of Fig. 1. These are defined formally as
the atomic position densities of states $P(z) = dN/dz$, where $dN$ is
the number of sp$^{2}$ or of sp$^{3}$ sites lying in the vertical position
interval between $z$ and $z+dz$.
The resulting distributions are shown in Fig. 2. There are two prominent
characteristics. The probability for sp$^{2}$ atoms at the surface layers
($\sim$ 2 \AA\ thickness) is pronounced in accordance with Fig. 1, while
the probability for sp$^{3}$ atoms in this layer is small. On the
other hand, at the middle of the slab where bulklike conditions
dominate, the probability for sp$^{3}$ atoms is maximized while for
sp$^{2}$ atoms is minimized. In general, there is an overall tendency 
through the whole slab to have enhancement of the probability for one
type of coordination in a certain region, and at the same time reduction
of the probability for the other type in this region.

\subsection{ {\it Atomic level stresses} }

Before addressing the dependence of local stresses on depth, we examine
their average distribution. As shown in Ref.\ \cite{Kelires1}, 
insight into the stress field is provided by calculating the probability 
distributions (stress density of states) over all atoms in the system.
We used the same analysis here for the cell discussed above.
The results are given in Fig. 3. $P(\sigma)$ is the unnormalized 
probability of finding an atom under local stress $\sigma$; positive
values indicate compressive stress. We show the decomposed
distributions according to coordination. The central conclusion from
this analysis is that all features observed previously in the bulk
system are still valid, despite the presence of large discontinuities
introduced by the surface. These features are: (a) The majority
of sp$^{3}$ atoms are under compressive stress with a peak (most probable)
value around $\sim$ 7 GPa. On the other hand, the majority
of sp$^{2}$ atoms are under tensile stress with a larger peak value around 
$\sim$ 20 GPa. This shows a tendency of the fewer sp$^{2}$ atoms to
compensate the compressive stress of the sp$^{3}$ atoms. (b) Both
distributions are broad covering compressive and tensile values,
so fourfold (threefold) atoms could be under tension (compression) as
well, although less likely. Thus, an {\it inhomogeneous stress 
distribution} arises in the network.

The stress inhomogeneity has another component in the present case:
the surface effect. To show quantitatively how the atomic stresses vary
as a function of depth, we compute the probability of finding a
threefold (fourfold) atom under tensile or compressive stress at
depth z. We plot in Fig. 4 the probabilities for threefold
and fourfold atoms to have tensile and compressive stress, respectively.
These are the dominant stress conditions for the two hybridizations,
as discussed above. It is clear that the probability of having tensile
stress at an sp$^{2}$ site is largest in the surface area, while it is
smaller deeper in the cell. To the contrary, the probability of having 
compressive stress at an sp$^{3}$ site is largest in the middle of
the slab where bulklike conditions dominate, and smaller at and near 
the surface. 

The stress probability is related to the position probability
shown in Fig. 2. A comparison of the two figures shows that where ever the
position probability of fourfold atoms is maximized, the probability
for tensile stress is minimized, and {\it vice versa}. This again  
indicates the relationship between bonding and local stress conditions.

\section{Discussion}

The present and previous \cite{Kelires1} results show
that the bonding hybridization - local stress relationship has a
global character. We would further like to comment on two
related issues.

It is stated, as a conclusion of thermal annealing experiments
\cite{Friedmann,Ferrari}, that although compressive stress is necessary
to sustain the sp$^{3}$ bonding in ta-C during deposition, it is not
so for post-growth processes where stress can be relieved without any
appreciable structural changes. This statement is valid when refering
to the average intrinsic stress of the film \cite{Kelires1}. However,
the local atomic stresses do not decrease drastically. They are only
re-arranged to yield a zero (small) net stress. In order to sustain
the sp$^{3}$ bonding during growth, and also during annealing, it is
necessary to preseve the local compressive conditions.

Since by definition the local atomic stress (hydrostatic pressure)
describes the local density fluctuations, it is logical to say that a
site under tension reflects a low local density. We know from the
above discussion that in such a site will most probably have
sp$^{2}$ bonding, but it is also quite possible for it to have sp$^{3}$
bonding. This means that there is not an absolute one-to-one
correspondence between bonding hybridization and local density.
Local densification \cite{Rober1} following ion subplantation 
(at optimum energies for small relaxation of the density increment)
might not actually lead to the formation of an sp$^{3}$ site. This
suggests that the formation of the sp$^{3}$-rich phase occurs
because the rate of densification-compressive stress-sp$^{3}$
bonding events overwhelms the rate of sp$^{2}$ bonding generation.

\section{Conclusions}

We examined in this paper the effect of the surface environment
on the local atomic stresses, which arise from incompatibilities
due to the non-equivalent bonding hybridizations in ta-C. 
We found a close relationship between hybridization and local stress
conditions, as previously found in simulations of the bulk system,
and we conclude that this relationship has a global character.

\begin{center}
\begin{figure}
\includegraphics[width=0.7\textwidth]{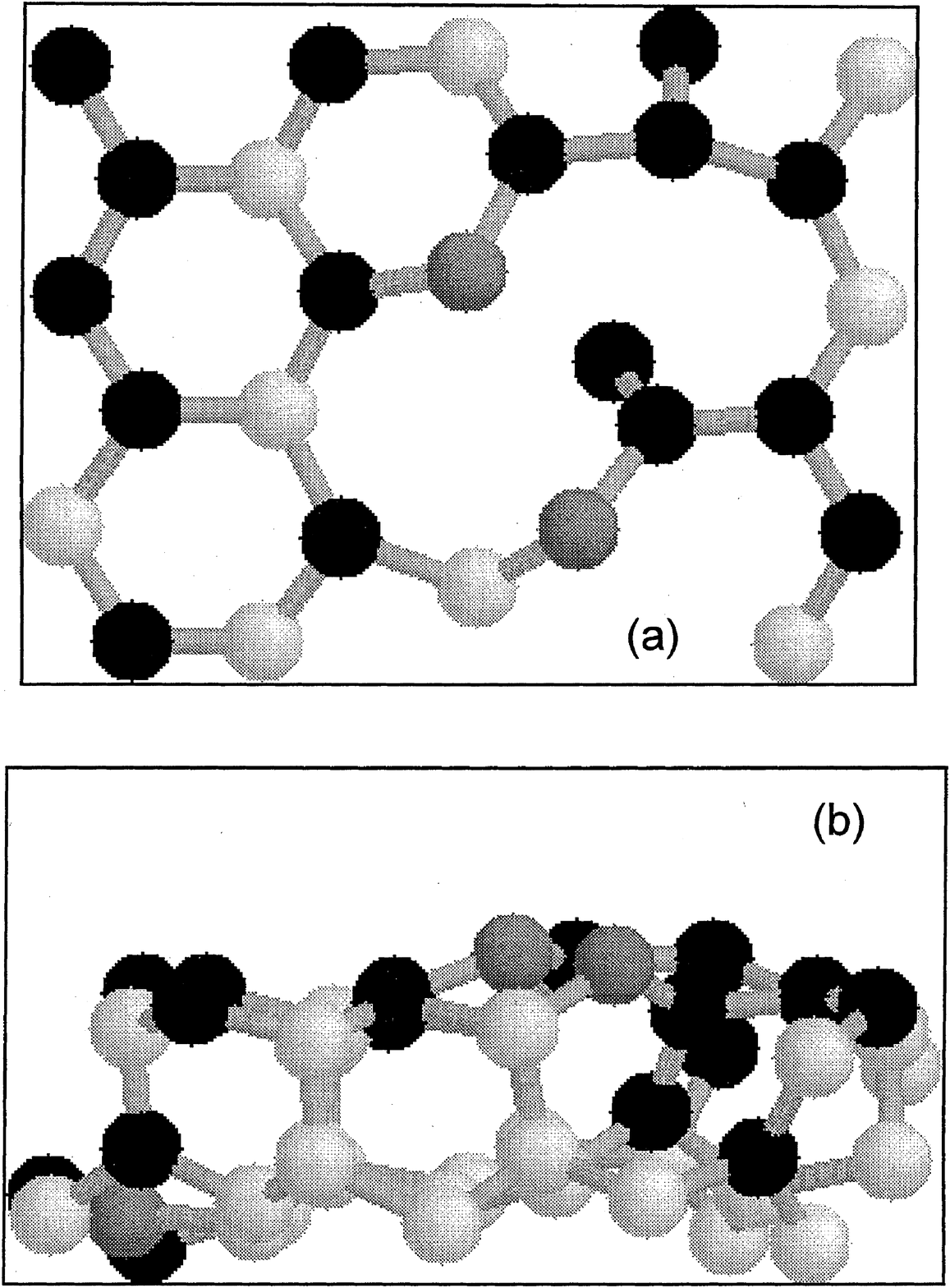}
\caption{Ball and stick model of the surface structure of a ta-C network
with $\bar{z} \simeq$ 3.75. (a) Top view showing the bonding in the
surface layer. (b) Side view showing the bonding of surface layer to
atoms beneath. Black balls stand for sp$^{2}$ atoms, white balls show
sp$^{3}$ atoms, and grey balls denote twofold atoms.}
\end{figure}

\vspace*{2cm}
\begin{figure}
\includegraphics[width=0.7\textwidth]{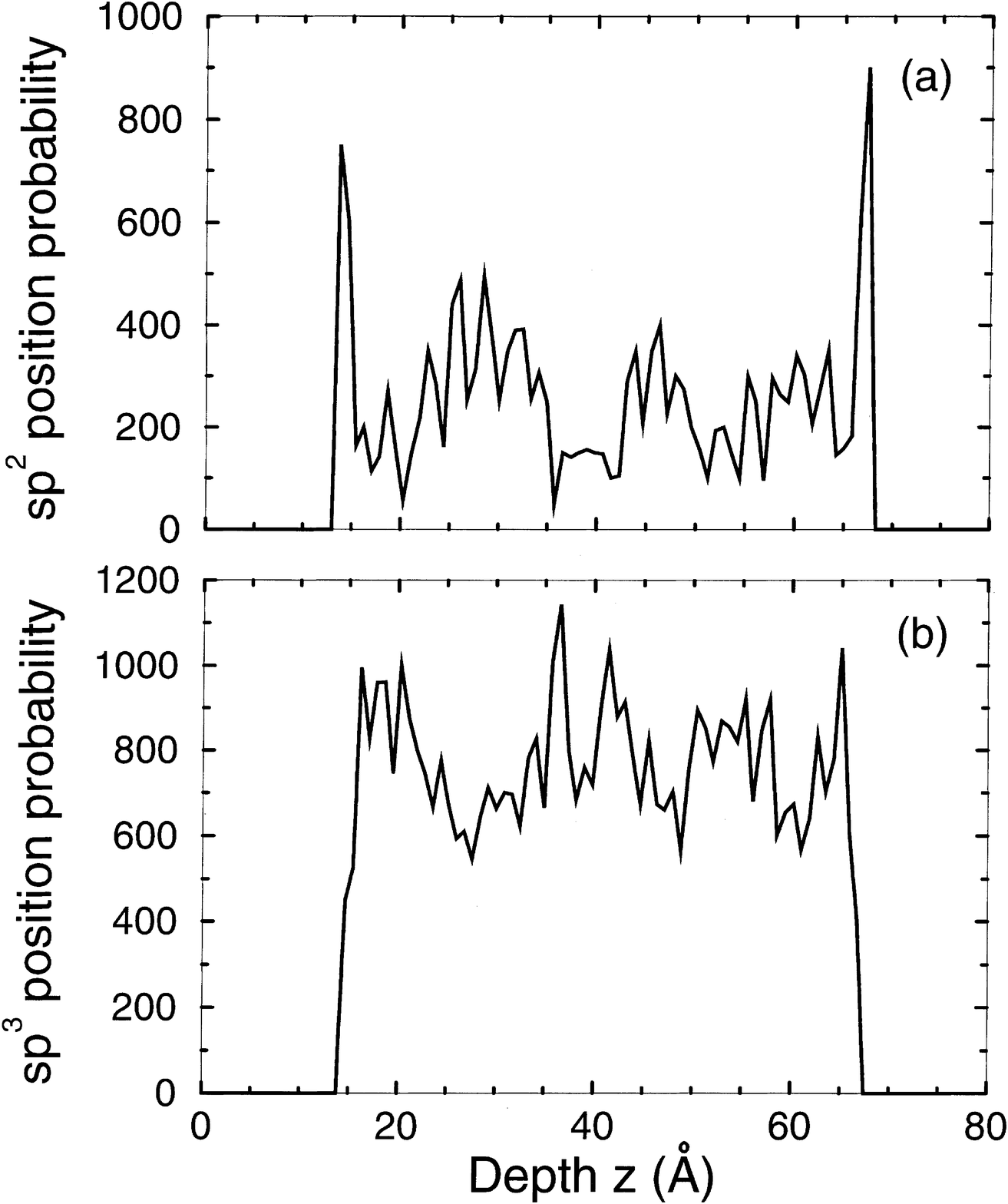}
\caption{Position probabilities vs. depth z, at 300 K, in the slab cell 
of Fig. 1. (a) For sp$^{2}$ atoms. (b) For sp$^{3}$ atoms.}
\end{figure}

\vspace*{5cm}
\begin{figure}
\includegraphics[width=0.7\textwidth]{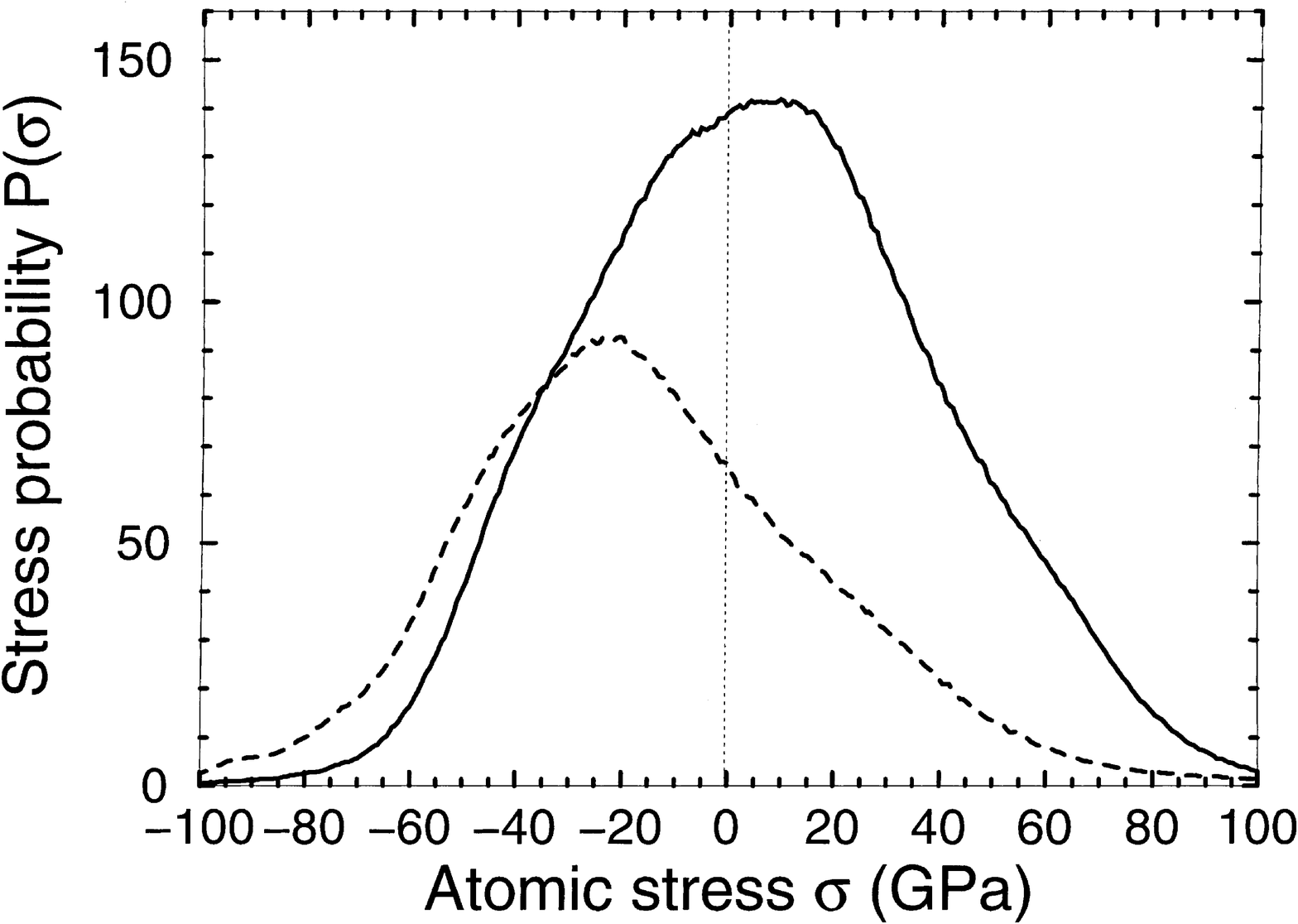}
\caption{Distributions of atomic stresses at 300 K in the slab cell 
of Fig. 1. Solid and dashed
lines are for sp$^{3}$ and sp$^{2}$ atoms, respectively.}
\end{figure}

\vspace*{2cm}

\begin{figure}
\includegraphics[width=0.7\textwidth]{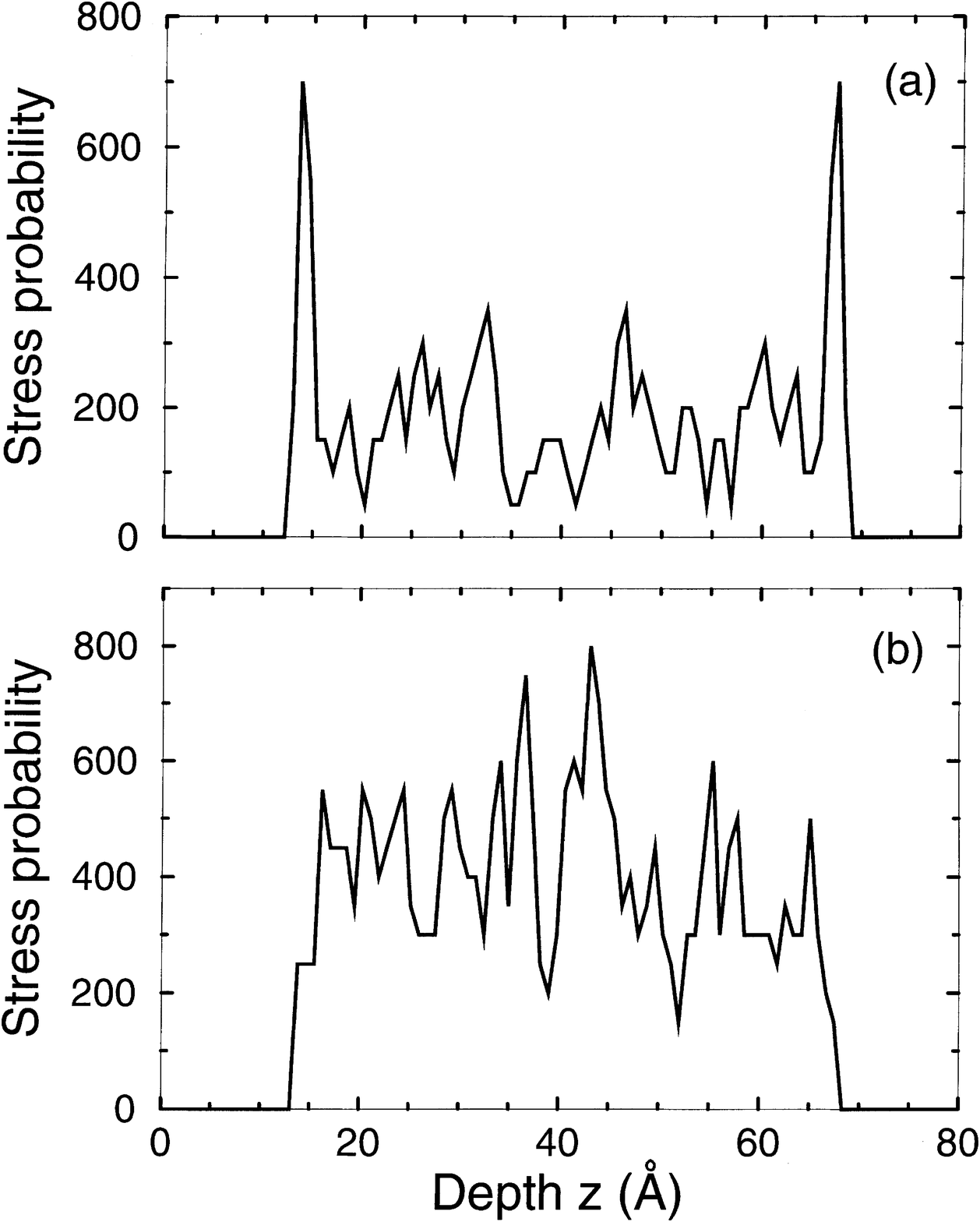}
\caption{Stress probability distributions vs. depth z, at 300 K, in 
the slab cell of Fig. 1. (a) For sp$^{2}$ atoms. (b) For sp$^{3}$ atoms.}
\end{figure}
\end{center}
\end{document}